\documentclass[aps,prd,nofootinbib,twocolumn,preprintnumbers]{revtex4}

\usepackage{amssymb}
\usepackage{amsmath}
\usepackage{epsfig}
\usepackage{hyperref}
\usepackage{url}
\usepackage{breakurl}
\usepackage{undertilde}
\usepackage{subfigure}

\makeatletter
\def\simgt{\mathrel{\lower2.5pt\vbox{\lineskip=0pt\baselineskip=0pt
           \hbox{$>$}\hbox{$\sim$}}}}
\def\simlt{\mathrel{\lower2.5pt\vbox{\lineskip=0pt\baselineskip=0pt
           \hbox{$<$}\hbox{$\sim$}}}}
\makeatother

\newcommand{\be}{\begin{equation}}
\newcommand{\ee}{\end{equation}}
\newcommand{\bea}{\begin{eqnarray}}
\newcommand{\eea}{\end{eqnarray}}
\newcommand{\Fig}[1]{Fig.~\ref{#1}}
\newcommand{\Eq}[1]{Eq.~(\ref{#1})}

\newcommand{\Sec}[1]{Sec.~\ref{#1}}

\newcommand{\vev}[1]{\langle #1 \rangle}

\newcommand{\LL}{\mathcal{L}}

\begin{document}

\preprint{UCB-PTH-11/12}

\title{Higgs Descendants}

\author{Clifford Cheung and Yasunori Nomura}
\affiliation{Berkeley Center for Theoretical Physics, 
  University of California, Berkeley, CA 94720, USA}
\affiliation{Theoretical Physics Group, 
  Lawrence Berkeley National Laboratory, Berkeley, CA 94720, USA}

\begin{abstract}

We define a Higgs descendant $\chi$ to be a particle beyond the standard 
model whose mass arises predominantly from the vacuum expectation value 
of the Higgs boson.  Higgs descendants arise naturally from new physics 
whose intrinsic mass scale is unrelated to the electroweak scale.  The 
coupling of $\chi$ to the Higgs boson is fixed by the mass and spin of 
$\chi$, yielding a highly predictive setup in which there may be substantial 
modifications to the properties of the Higgs boson.  For example, if the 
decay of the Higgs boson to $\chi$ is kinematically allowed, then this 
branching ratio is largely determined.  Depending on the stability of 
$\chi$, Higgs decays may result in a variety of possible visible or 
invisible final states.  Alternatively, loops of $\chi$ may affect Higgs 
boson production or its decays to standard model particles.  If $\chi$ 
is stable dark matter, then the mandatory coupling between $\chi$ and 
the Higgs boson gives a lower bound on the direct detection cross section 
as a function of the $\chi$ mass.  We also present a number of explicit 
models which are examples of Higgs descendants.  Finally, we comment on 
Higgs descendants in the context of the excesses near $125~{\rm GeV}$ 
recently observed at ATLAS and CMS.

\end{abstract}

\maketitle

\section{Introduction}
\label{sec:intro}

A primary goal of the LHC is to reveal the fundamental dynamics underlying 
electroweak symmetry breaking.  The characteristic mass scale of the 
standard model is the vacuum expectation value (VEV) of the Higgs field, 
$v = 246~{\rm GeV}$, which is radiatively unstable.  This fact has been 
the central driving force for exploring new theories beyond the standard 
model at the TeV scale.

Theories which address the electroweak hierarchy problem universally 
introduce dimensionful parameters from which $v$ originates.  For example, 
in the case of supersymmetry~\cite{Witten:1981nf} the $\mu$ term and soft 
masses determine the value of $v$.  Likewise, in theories with a pseudo 
Nambu-Goldstone Higgs boson~\cite{Kaplan:1983fs}, the scale $v$ descends 
a loop down from the symmetry breaking decay constant $f$.  In such 
instances the proximity of new mass scales to $v$ is {\it required}, 
and thus not a coincidence.

Conversely, for any new physics unrelated to the origin of electroweak 
symmetry breaking, associated dimensionful parameters have, a priori, 
no reason to be near $v$.  If these mass scales lie far above the 
electroweak scale, then the associated particles are kinematically 
inaccessible to experiments like the LHC, and we can integrate them out. 
Hence, in this limit our only hope of probing such new physics is from 
higher dimension operators, e.g.\ as in the case of the Weinberg operator 
for small neutrino masses~\cite{Weinberg:1980bf}.

If, on the other hand, the new mass scales are very tiny then the 
associated states will be, naively, quite light.  This naive intuition, 
however, fails if the new physics is coupled to the standard model.  In 
this case there will in general be induced interactions between these 
new states and the Higgs field.  After electroweak symmetry breaking, 
such couplings provide the main contribution to the masses of these new 
states, which are necessarily proportional to some power of $v$.  We 
will refer to this broad class of new states as Higgs descendants, and 
as we will see they can have a drastic impact on Higgs physics even 
though they do little to address the hierarchy problem.

More concretely, let us define a Higgs descendant as an additional 
particle beyond the standard model, $\chi$, whose mass obeys the property 
that
\bea
  v \rightarrow 0 \quad &\Rightarrow& \quad m_\chi \rightarrow 0.
\label{eq:definition}
\eea
Note that this condition does not hold typically for new states associated 
with the sectors addressing the hierarchy problem.  For instance, in 
supersymmetry the superpartner masses arise from soft supersymmetry 
breaking parameters, so they do not necessarily vanish as $v \rightarrow 0$. 
Likewise, any particle with a bare mass is not a Higgs descendant.

Still, Higgs descendants represent large classes of theories that may appear 
in varying contexts.  Several examples of Higgs descendants can already 
be found in the literature.  For instance, a number of authors have 
considered minimal singlet dark matter coupled via the so-called Higgs 
portal~\cite{Silveira:1985rk,Baek:2011eg}.  In many cases, the singlet 
is given an intrinsic mass scale which may be fixed by other considerations, 
e.g.~a thermal relic abundance \cite{Raidal:2011xk}.  Restricting such 
theories to forbid/suppress explicit mass terms for the dark matter 
yields a Higgs descendant~\cite{McDonald:1993ex}.  Likewise, fourth 
generation Dirac neutrinos~\cite{Ishiwata:2011hr} and certain hidden 
sector models~\cite{Baumgart:2009tn,Cheung:2010jx} are also examples 
of Higgs descendants.

The condition of \Eq{eq:definition} implies that $m_\chi$ can be expanded 
in powers of $v$, so for example
\be
  m_\chi = \lambda v^n + \ldots,
\label{eq:taylor-f}
\ee
for a fermionic Higgs descendant $\chi$.  Here, $n > 0$, and $\lambda$ 
parametrizes our ignorance of physics coupling the Higgs field to the 
Higgs descendant.  For weakly coupled theories, $n$ is integer, but 
in general it need not be \cite{Stancato:2008mp}.  The ellipses denote 
higher order terms in $v$, which we will ignore throughout.  Higgs 
descendant theories possess a highly restricted phenomenology because 
the interactions relevant to many physical processes are essentially 
fixed by $m_\chi$ alone.  In particular, by replacing the Higgs VEV as
\be
  v \rightarrow v+ h,
\ee
where $h$ is the propagating Higgs boson field, \Eq{eq:taylor-f} leads 
to a Lagrangian of the form
\be
  \LL = -m_\chi \left( 1+ \frac{n h}{v}\right) \bar\chi \chi +\ldots ,
\label{eq:hcoupling}
\ee
where the ellipses denote terms higher order in the Higgs field and 
we have considered a Dirac fermion $\chi$ for concreteness.

For a scalar Higgs descendant $\chi$, we define the index $n$ according to
\be
  m_\chi^2 = \lambda v^n + \ldots,
\label{eq:taylor-s}
\ee
reflecting the dimension of the $\chi$ mass term in the Lagrangian. 
Equation~(\ref{eq:hcoupling}) then applies equally for a complex scalar 
$\chi$ only by making $m_\chi \rightarrow m_\chi^2$.  The generalization 
is obvious for other spins and real representations.  Note that throughout, 
$n$ is defined by \Eq{eq:taylor-f} and \Eq{eq:taylor-s} for fermionic 
and bosonic $\chi$, respectively.

The organization of our paper is as follows.  In Section~\ref{sec:Higgs-prop}, 
we discuss how a Higgs descendant can have significant implications for 
Higgs phenomenology at the LHC.  These arise either as new decay modes or 
as modifications of the standard model production and decay modes of the 
Higgs boson.  We discuss the case in which the Higgs descendants are stable 
or unstable decaying into standard model particles, leading to distinct 
signatures for Higgs boson decays.  In Section~\ref{sec:DM}, we consider 
the case in which a Higgs descendant is stable and comprises the dark matter 
of the universe.  In this case, \Eq{eq:hcoupling} provides a lower bound, 
modulo unnatural cancellations, on the dark matter-nucleon scattering 
cross section relevant for direct detection experiments, as a function 
of $m_\chi$.  Correlations between dark matter and LHC physics would then provide 
a powerful probe of the underlying theory.  In Section~\ref{sec:models}, 
we present simple, explicit models of Higgs descendants.  Finally, we 
conclude in Section~\ref{sec:discuss}.

\section{Higgs Boson Properties}
\label{sec:Higgs-prop}

The existence of a Higgs descendant $\chi$ leads to the mandatory coupling 
in \Eq{eq:hcoupling}.  Here we discuss the effect of this coupling on 
Higgs physics.  Throughout this paper, we will limit ourselves to the case 
of a single Higgs doublet.

\subsection{Light Higgs Descendants}
\label{subsec:h_chi-chi}

Consider a scenario of light Higgs descendants, defined as $m_\chi < m_h/2$. 
In this case the Higgs boson has a new decay channel, $h \rightarrow 
\chi\chi$, with a branching ratio fixed by $m_h$, $m_\chi$, and $n$, 
which is independent of the gauge quantum numbers of $\chi$ (modulo 
multiplicity factors).%
\footnote{If $n$ is odd, it might naively be thought that $\chi$ must 
 carry standard model charges, but this is not true.  The mass of $\chi$ 
 may arise from a singlet field $s$ whose VEV is induced by the Higgs 
 VEV, i.e.~$\vev{s} \propto v$.}
 
The partial decay rate of the Higgs boson to standard model particles can 
be found in Ref.~\cite{Djouadi:2005gi}, while the partial decay rate to 
$\chi$ is given by
\be
  \Gamma_f = \frac{m_h}{8\pi} 
    \biggl( \frac{n m_\chi}{v} \biggr)^2 
    \Biggl( {1- \frac{4m_\chi^2}{m_h^2}} \Biggr)^{3/2},
\label{eq:chi-Dirac}
\ee
for a Dirac fermion $\chi$ and by
\be
  \Gamma_s = \frac{m_h}{16\pi} 
    \biggl( \frac{m_\chi}{m_h} \biggr)^2 \biggl( \frac{n m_\chi}{v} \biggr)^2 
    \Biggl( {1- \frac{4m_\chi^2}{m_h^2}} \Biggr)^{1/2},
\label{eq:chi-cscalar}
\ee
for a complex scalar $\chi$.  For a Majorana fermion or a real scalar, 
one makes the replacements $\Gamma_f \rightarrow \Gamma_f /2$ and 
$\Gamma_s \rightarrow \Gamma_s /2$, respectively.  The additional factor 
of $m_\chi^2 /m_h^2$ in \Eq{eq:chi-cscalar} relative to \Eq{eq:chi-Dirac} 
will produce quantitatively different physics for fermionic versus scalar 
$\chi$, especially for small $m_\chi$.

\begin{figure*}[t]
\centering
 \subfigure{\includegraphics[scale=0.9]{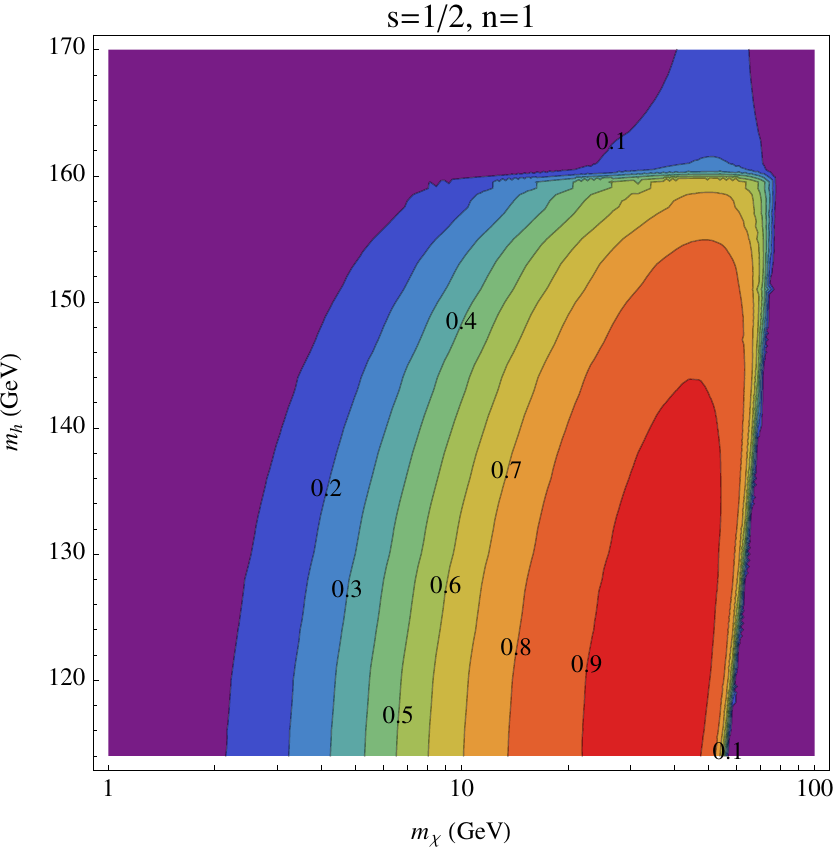}}
\hspace{1cm}
 \subfigure{\includegraphics[scale=0.9]{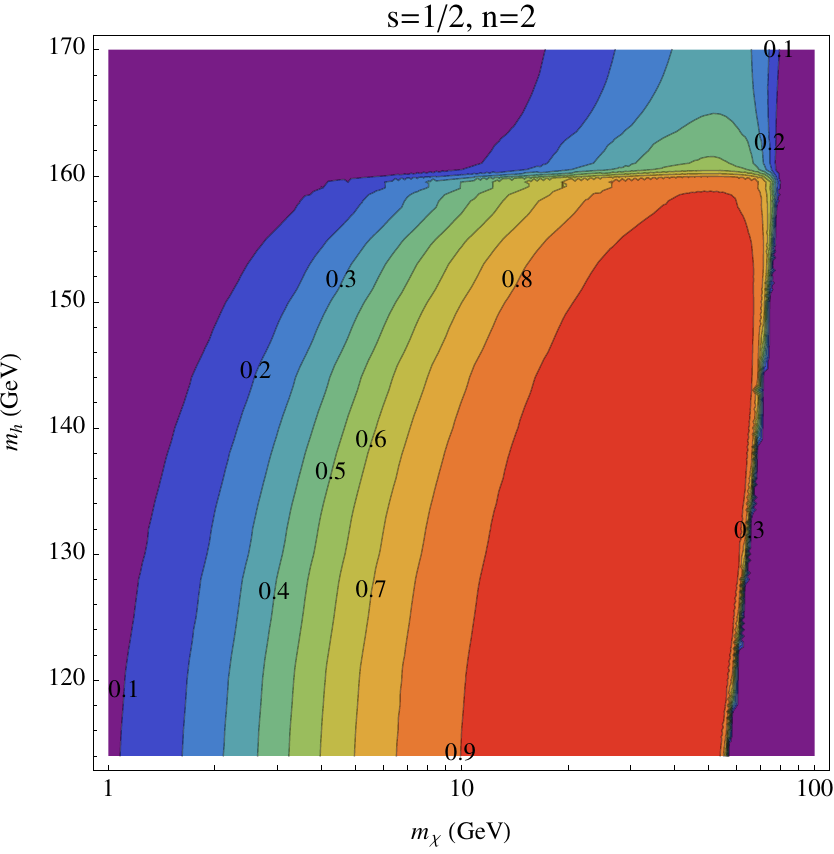}}
\caption{Branching ratio of $h \rightarrow \chi\chi$ for a (Dirac) 
 fermionic $\chi$ with $n=1$ (left) and $n=2$ (right).}
\label{fig:fermion}
\end{figure*}

The branching ratio ${\rm BR}(h \rightarrow \chi\chi)$ is shown in 
\Fig{fig:fermion}  for a Dirac fermion $\chi$ as a function of $m_\chi$ 
and $m_h$ for $n=1$ and $2$.
One expects that for $m_h \simlt 2 m_W$, the Higgs branching ratio 
into a Higgs descendant becomes important when $n m_\chi$ is of order 
$m_b$, the bottom quark mass, and this is indeed the case.  For example, 
${\rm BR}(h \rightarrow \chi\chi) \gtrsim 50\%$ for a light Higgs 
boson when $m_\chi \simgt 6~{\rm GeV}$ ($m_\chi \simgt 3~{\rm GeV}$) 
for $n=1$ ($n=2$).  For a heavy Higgs boson above the $WW$ threshold, 
decays into the Higgs descendant can be non-negligible only in small 
regions of parameter space.

\begin{figure*}[t]
\centering
 \subfigure{\includegraphics[scale=0.9]{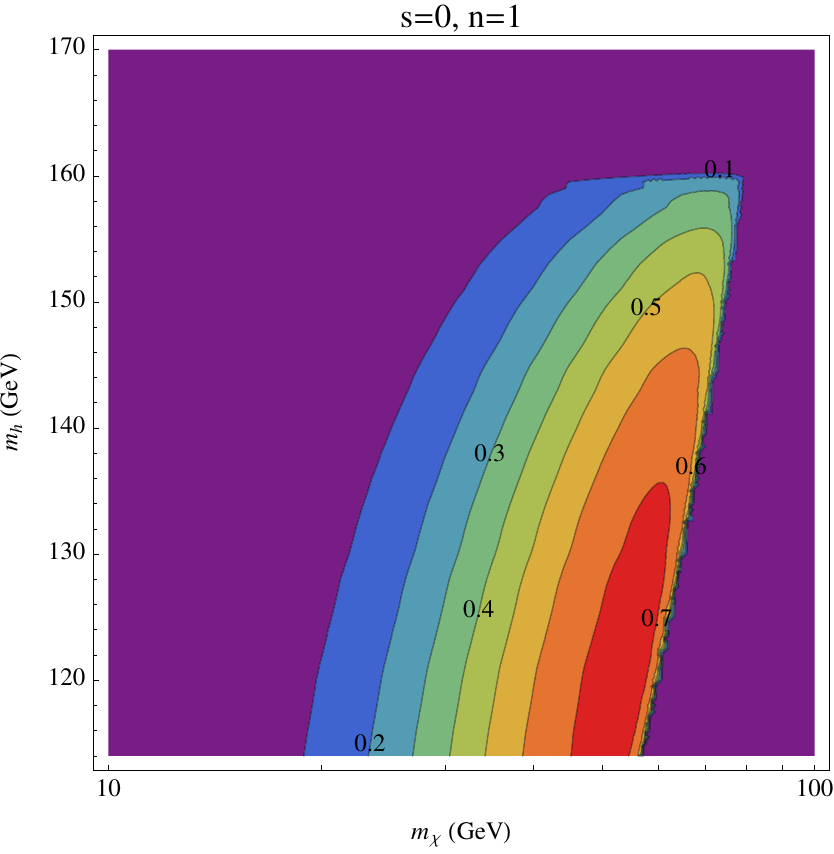}}
\hspace{1cm}
 \subfigure{\includegraphics[scale=0.9]{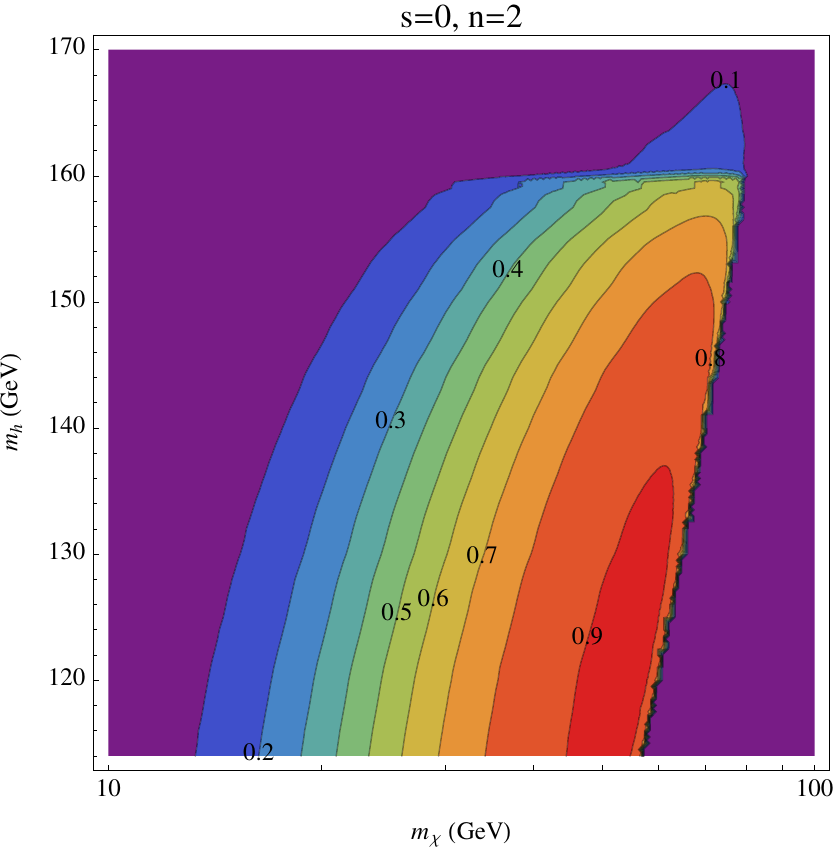}}
\caption{Branching ratio of $h \rightarrow \chi\chi$ for a (complex) 
 scalar $\chi$ with $n=1$ (left) and $n=2$ (right).}
\label{fig:scalar}
\end{figure*}

If $\chi$ is a scalar, then the story changes quantitatively.  In this 
case, $\chi$ interacts with the Higgs boson with a dimensionful coupling 
$n m_\chi^2/v$, rather than $n m_\chi/v$.  This induces an additional 
factor of $m_\chi^2/m_h^2$ which suppresses the Higgs boson decay rate 
to Higgs descendants, especially for small $m_\chi$.  This trend is 
verified in \Fig{fig:scalar}, which depicts ${\rm BR}(h \rightarrow 
\chi\chi)$ for complex scalar $\chi$.
We find that ${\rm BR}(h \rightarrow \chi\chi) \gtrsim 50\%$ for a light 
Higgs boson only when $m_\chi \simgt 30~{\rm GeV}$ ($m_\chi \simgt 
20~{\rm GeV}$) for $n=1$ ($n=2$).  Decays into scalar Higgs descendants 
never dominate for a heavy Higgs boson above the $WW$ threshold.

What occurs after the Higgs decays to a pair of $\chi$ particles is 
somewhat more model dependent.  Hence, the resulting experimental signatures 
are as well.  The interactions in \Eq{eq:hcoupling} preserve a $\mathbb{Z}_2$ 
symmetry which, if unbroken, ensures the stability of $\chi$.  Consider 
first the case where this $\mathbb{Z}_2$ is sufficiently exact that $\chi$ 
is stable on collider time scales.  In this case, the signatures depend 
largely on the standard model gauge quantum numbers of $\chi$, which we 
now consider cases by case.

\subsubsection{Stable Case}

If $\chi$ is colored it will hadronize, yielding highly ionizing, sometimes 
intermittent tracks caused by long-lived, charged hadronic states.%
\footnote{For discussions of such signatures in other contexts, see 
 e.g.\ Ref.~\cite{Barbieri:2000vh}.  Note also that the Higgs decay 
 widths in this case are $N$ times those in Eqs.~(\ref{eq:chi-Dirac}) 
 and (\ref{eq:chi-cscalar}), where $N = 3$ and $8$ for a color triplet 
 and octet $\chi$, respectively.}
This possibility, however, is strongly constrained by direct production 
of $\chi$ at the LHC~\cite{Chatrchyan:2012sp}, and is not viable unless 
the hadronization leads mostly to neutral bound states.

Next, let us consider the scenario in which the Higgs descendant is 
uncolored, but carries electroweak charge.  For example, if the Higgs 
descendant has the quantum numbers of a lepton, then $\chi = (\chi^0, 
\chi^+)$.  In this case the Higgs boson may decay into the charged 
components $\chi^+$, possibly as well as the neutral components 
$\chi^0$, depending on the quantum numbers of the $\chi$/$\bar{\chi}$ 
multiplets.  The Higgs descendant then may lead to signatures discussed 
in Ref.~\cite{Ibe:2006de} if the mass splitting $\varDelta m \equiv 
m_{\chi^+} - m_{\chi^0}$ is sufficiently small---for $\varDelta m 
\simlt m_\pi$ it leads to stable charged tracks, while for $m_\pi 
\simlt \varDelta m \simlt 200~{\rm MeV}$ to short tracks of $\chi^+$ 
together with soft pions from $\chi^+ \rightarrow \chi^0 \pi^+$.  If 
$\varDelta m$ is larger (but not much larger than $\simeq {\rm GeV}$), 
then both $h \rightarrow \chi^+ \chi^-$ and $h \rightarrow \chi^0 \chi^0$ 
are recognized only as invisible Higgs boson decays at the LHC (unless 
boost factors for $\chi^+$ are large due to $m_\chi \ll m_h$), with 
the widths larger than those in Eqs.~(\ref{eq:chi-Dirac}) and 
(\ref{eq:chi-cscalar}) due to an appropriate multiplicity factor. 
For $\varDelta m > {\rm GeV}$, decay products of $\chi^+$ may be 
directly tagged.

Finally, if $\chi$ is not charged under any standard model gauge 
interactions, then $h \rightarrow \chi\chi$ contributes to the invisible 
Higgs decay width.  (For recent discussions on invisibly decaying Higgs 
bosons, see e.g.\ Ref.~\cite{Bai:2011kq}.)  If exactly stable and neutral, 
$\chi$ is a possible dark matter candidate, which we discuss in depth 
in \Sec{sec:DM}.

\subsubsection{Unstable Case}

In general, the $\mathbb{Z}_2$ symmetry preserved by \Eq{eq:hcoupling} may 
be broken if there are additional couplings of $\chi$ to standard model 
particles.  These interactions permit the Higgs descendant to decay 
visibly inside the collider, as discussed in a similarly general context 
in Ref.~\cite{Englert:2011us}.  For example, for a singlet fermionic $\chi$, 
consider the interaction $\chi {\cal O}$, where
\be
  {\cal O} = \{ u^c d^c d^c,  \ell \ell e^c, q \ell d^c, q \ell^\dagger u^c, 
    u^c d^{c\dagger} e^c ,\ell h \}.
\ee
Here, ${\cal O}$ is constructed from all leading singlet fermionic standard 
model composites.  Each of these operators affects the phenomenology in 
a distinct way.

For ${\cal O}= u^c d^c d^c$ each Higgs descendant decays to three jets, 
yielding a striking six jet final state from Higgs boson decays.  In this 
scenario, decays of the Higgs boson may be buried in soft jets and thus 
difficult to discern at the LHC.  This operator also induces low energy 
mixing between $\chi$ and standard model baryons.  For ${\cal O} = \ell 
\ell e^c$, the Higgs descendant decays via di-lepton plus missing energy. 
For ${\cal O} = q \ell d^c, q \ell^\dagger u^c, u^c d^{c\dagger}e^c$, 
each Higgs descendant decays to di-jet plus lepton (or di-jet plus missing 
energy for the first two operators).  For ${\cal O} = \ell h$, the Higgs 
descendant decays to di-jet or di-lepton plus missing energy, or di-jet 
plus lepton.  However, this operator is not ideal because it generates 
a large Dirac mass between $\chi$ and neutrinos, which is difficult to 
reconcile with neutrino oscillation experiments and cosmological constraints.

Lastly, consider the case of a singlet scalar $\chi$, which can couple 
via $\chi {\cal O}$ where
\be
  {\cal O} = \{ q h u^c, q h^\dagger d^c, \ell h^\dagger e^c, 
    F_{\mu\nu}F^{\mu\nu}, G_{\mu\nu}G^{\mu\nu} \}.
\ee
Depending on the operator, $\chi$ will decay to di-jet, di-lepton, 
or di-gamma.

It is straightforward to apply similar arguments to $\chi$ which is 
not a standard model singlet, but we will not do so here.

\subsection{Heavy Higgs Descendants}
\label{subsec:modified}

Let us now consider the case of heavy Higgs descendants, where 
$m_\chi > m_h/2$.  In this case, the Higgs boson is kinematically 
forbidden from decaying to Higgs descendants.  One might think naively 
that the properties of the Higgs boson are very similar to that of 
the standard model.  However, in certain cases $\chi$ is charged 
under the standard model gauge group, in which case it will typically 
influence Higgs boson production via $gg \rightarrow h$ as well as 
the decay $h \rightarrow \gamma\gamma$.  Moreover, as we will see, 
the contributions to these processes from Higgs descendants are fixed 
by the choice of $n$ and the spin and charges of $\chi$.

The coupling of the Higgs boson to the gluon is
\bea
  {\cal L}_{hgg} &=& \frac{\alpha_s}{12\pi}\frac{h}{v} 
    \left( \sum_i c_i t_i N_i \frac{\partial \log m_i(v)}{\partial \log v} \right) 
    G^a_{\mu\nu} G^{a\mu\nu}
\nonumber\\
  &=& \frac{\alpha_s}{12\pi}\frac{h}{v} 
    \Bigl( 1 + c_\chi t_\chi N_\chi n \Bigr) G^a_{\mu\nu} G^{a\mu\nu},
\label{eq:hgg}
\eea
where $i$ labels heavy species which provide threshold corrections to the 
beta function of QCD.  Here $c_i = 2$ for Dirac fermions and $c_i = 1/2$ for 
complex scalars, $t_i$ is the Dynkin index of the multiplet, and $N_i$ is 
the multiplicity.  In the second line we have plugged in for $i=\chi$ and 
the top quark, where $\chi$ is colored.  We see that the Higgs boson coupling 
to the gluon is fixed by a set of discrete quantum numbers of the $\chi$ 
field.  For example, the fourth generation quarks have $\{c_\chi, t_\chi, 
N_\chi, n\} = \{2,1/2,2,1\}$, enhancing the amplitude for Higgs production 
through gluon fusion by a factor of three and thus the production 
cross section by a factor of nine.

A similar formula for the coupling of Higgs boson to the photon can be 
derived from the beta function of QED:
\bea
  {\cal L}_{h\gamma\gamma} &=& \frac{\alpha}{12\pi} \frac{h}{v} 
    \left( -\frac{21}{2} + \sum_i c_i q_i^2 N_i 
    \frac{\partial \log m_i(v)}{\partial \log v} \right) F_{\mu\nu} F^{\mu\nu}
\nonumber\\
  &=& \frac{\alpha}{12\pi} \frac{h}{v} 
    \left( -\frac{21}{2} + \frac{8}{3} + c_\chi q_\chi^2 N_\chi n \right) 
    F_{\mu\nu} F^{\mu\nu},
\label{eq:hgaga}
\eea
where the Dynkin index $t_i$ is replaced with the electric charge $q_i^2$, 
and the term of $-21/2$ is the contribution to this coupling from a $W$ 
boson loop.  As before, we have plugged in for $i=\chi$ and the top quark 
in the second line, assuming an electrically charged Higgs descendant. 
As is well-known, the top quark and $W$ loop contributions to the Higgs 
coupling to photons destructively interfere~\cite{Djouadi:2005gi}. 
Similarly, loops of Higgs descendants also cancel against the $W$ loop 
contribution, since the mass of a Higgs descendant always grows with 
the Higgs VEV.  For example, this occurs for a fourth generation lepton, 
which contributes $\{c_\chi, q_\chi, N_\chi, n\} = \{2,1,1,1\}$.

The defining relation of our scenario, \Eq{eq:taylor-f}, implies that 
the mass of a Higgs descendant cannot be arbitrarily large, $m_\chi 
\lesssim 1~{\rm TeV}$, so it is subject to direct search limits from 
the LHC.  We now consider those limits.  While there are no dedicated 
LHC searches for states with arbitrary charges, we can get rough estimates 
for the limits from similar searches.

If $\chi$ is long-lived on collider time scales, then LHC searches for 
stable colored or electrically charged particles apply.  These mass 
limits are quite stringent for gluino-like ($\gtrsim 1~{\rm TeV}$) and 
squark-like ($\gtrsim 700~{\rm GeV}$) states, but relatively weak for 
slepton-like ($\gtrsim 200~{\rm GeV}$) states~\cite{Chatrchyan:2012sp}. 
We can thus conclude that long-lived Higgs descendants are substantially 
constrained if colored, but are quite viable otherwise.

On the other hand, if $\chi$ decays promptly then limits depend sensitively 
on its charges and decay modes.  For instance, consider $\chi$ which 
is colored and promptly decays.  For $\chi$ decaying to standard model 
quarks plus one or more leptons, the decay topology and thus the LHC 
limit are similar to that of top and bottom partner quarks ($\gtrsim 
600~{\rm GeV}$)~\cite{CMS:2012ab}.  Meanwhile, for $\chi$ decaying to 
standard model quarks plus missing energy, supersymmetry searches apply. 
Using the simplified model analysis of Ref.~\cite{Simplified}, one 
can estimate the LHC limit for a single squark-like state ($\gtrsim 
300~{\rm GeV}$), although this bound can be eliminated altogether if 
the mass of the missing energy particle is sufficiently large.  If, on 
the other hand, $\chi$ decays entirely to jets with no missing energy, 
as is the case in many $R$-parity violating models, bounds will be 
substantially weaker.

Next, let us consider the case of color-neutral, electroweak charged 
$\chi$ which promptly decays.  In this scenario, limits are far weaker 
due to the smaller production cross section.  For chargino-like states, 
one can apply constraints from LHC supersymmetry searches ($\gtrsim 
300~{\rm GeV}$)~\cite{Aad:2012cw} although these bounds make very 
specific assumptions about the cascade decay---in particular these 
searches are driven by the presence of intermediate sleptons which 
provide additional leptons.  In general, the limits on the minimal 
chargino-neutralino system are very weak at LHC, and the dominant bounds 
still come from LEP ($\gtrsim 100~{\rm GeV}$).

Let us now determine the effect of \Eq{eq:hgg} and \Eq{eq:hgaga} on Higgs 
boson phenomenology.  To do so we define a function $r[x]$ to be the ratio 
of a quantity $x$ in a particular theory divided by the value of $x$ in 
the standard model.
\begin{figure}[t]
\begin{center}
  \includegraphics[scale=0.9]{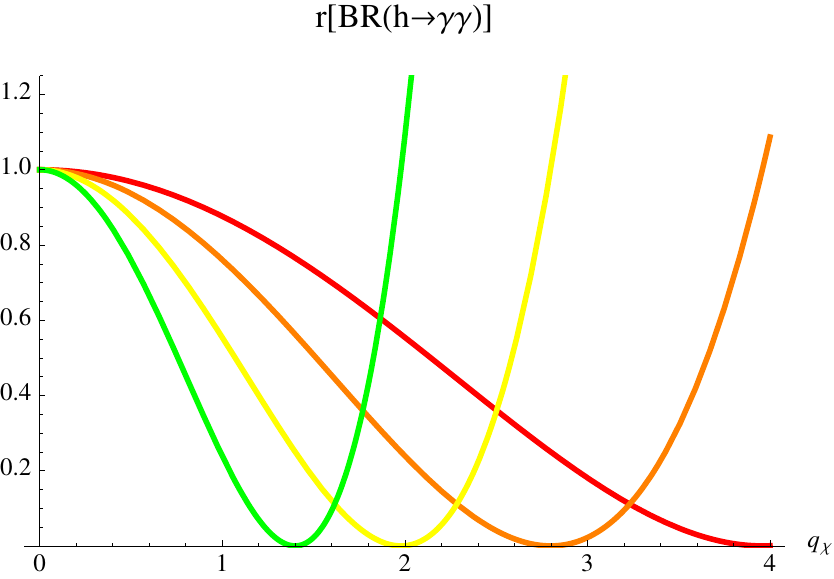}
\end{center}
\caption{Ratio of ${\rm BR}(h \rightarrow \gamma \gamma)$ versus its 
 standard model value for a color singlet $\chi$ of electric charge 
 $q_\chi$.  Here, \{red, orange\} correspond to a complex scalar with 
 $n = \{1,2\}$ and \{yellow, green\} correspond to a Dirac fermion 
 with $n = \{1,2\}$.}
\label{fig:BRdigamma}
\end{figure}
\begin{figure}[t]
\begin{center} 
  \includegraphics[scale=0.9]{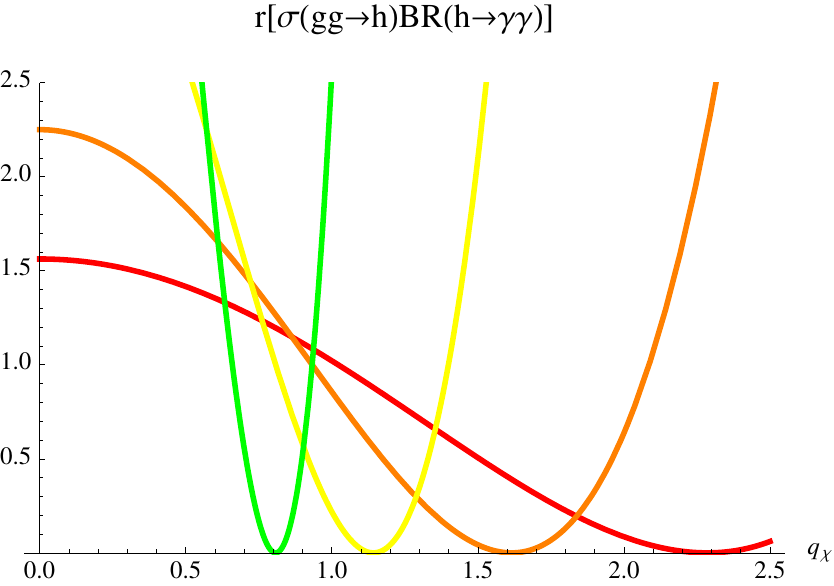}
\end{center}
\caption{Ratio of $\sigma(gg \rightarrow h) {\rm BR}(h \rightarrow 
 \gamma \gamma)$ versus its standard model value for a color 
 triplet $\chi$ of electric charge $q_\chi$.  The colors are 
 as in \Fig{fig:BRdigamma}.}
\label{fig:sBRdigamma}
\end{figure}
In \Fig{fig:BRdigamma} we have plotted $r[{\rm BR}(h \rightarrow \gamma 
\gamma)]$ for a color singlet $\chi$ with electric charge equal to $q_\chi$. 
As expected, the branching ratio to di-gamma diminishes for small values 
of $q_\chi$ due to the destructive interference with the $W$ boson. 
For sufficiently large $q_\chi$, however, the contribution from $\chi$ 
dominates and the di-gamma branching ratio begins to increase.  As 
discussed, a color singlet $\chi$ is relatively unconstrained by direct 
search limits and can be coupled perturbatively to the Higgs boson. 
In \Fig{fig:sBRdigamma}, we have plotted $r[\sigma(gg \rightarrow h) 
{\rm BR}(h \rightarrow \gamma \gamma)]$ for a color triplet $\chi$ 
with electric charge $q_\chi$.  Here the rate of di-gamma events from 
Higgs boson decays is substantially increased at low $q_\chi$, but 
can again cancel at larger values.  We emphasize, however, that because 
of direct search limits on colored particles, much of this parameter 
space is disfavored, without hiding $\chi$ by making it decay into 
multiple jets or invoking non-perturbative dynamics to sufficiently 
lift the mass of $\chi$.

\section{Higgs Descendant Dark Matter}
\label{sec:DM}

If a Higgs descendant $\chi$ contains a neutral and stable component, 
then this can comprise the dark matter of the universe.  In this section, 
we consider this possibility, especially its implications for direct 
detection experiments.  In what follows, we do not impose any constraints 
from demanding a thermal relic abundance of $\chi$ particles through 
freeze-out through the interaction in \Eq{eq:hcoupling}.  As noted in 
Ref.~\cite{Pospelov:2011yp}, for example, there is much leeway in 
generating dark matter of this type through non-thermal methods.

As a consequence of \Eq{eq:hcoupling}, $\chi$ has a mandatory coupling 
to the Higgs boson.  Since the Higgs boson 
couples to quarks and gluons, one in turn should expect an irreducible 
spin-independent scattering cross section of $\chi$ against a target 
nucleus via $t$-channel Higgs boson exchange.  For a Dirac fermion $\chi$, 
the cross section is
\be
  \sigma = \frac{\mu^2}{\pi} \left( \frac{n m_{\chi}}{v} \right)^2 
    \frac{1}{m_h^4} \left( Z g_{hpp} + (A-Z)g_{hnn} \right)^2,
\ee
where $\mu$ is the dark matter-nucleus reduced mass, $m_h$ is the Higgs 
boson mass, $Z$ and $A$ are the atomic number and weight of the target 
nucleus, and $g_{hNN}$ for $N = p,n$ are the coupling of the Higgs boson 
to the proton and neutron:
\be
  g_{hNN} = \frac{m_N}{ v} \sum_{q} f^N_q.
\label{eq:g-sNN}
\ee
The scattering cross section for complex scalar $\chi$ is one quarter 
that of a Dirac fermion.

To derive the dark matter-nucleon spin-independent cross section 
$\sigma_{\rm SI}$, we set $A = Z = 1$ and $\mu = 1~{\rm GeV}$.  Using 
$\sum_q f^N_q \simeq 0.30 \pm 0.015$, compiled in Ref.~\cite{Raidal:2011xk} 
based on Ref.~\cite{Giedt:2009mr}, we obtain
\be
  \sigma_{\rm SI} \simeq 2.5 \times 10^{-45}~{\rm cm}^2 
    \left( \frac{n m_\chi}{10~{\rm GeV}} \right)^2 
    \left(\frac{114~{\rm GeV}}{m_h} \right)^4,
\label{eq:sigma}
\ee
for Dirac fermion dark matter.  This gives a lower bound on the dark matter 
detection cross section; unless there is a substantial cancellation between 
the Higgs-exchange and other processes, we expect that the true dark 
matter-nucleon cross section is comparable to or larger than the value 
given in \Eq{eq:sigma}.

\begin{figure}[t]
\begin{center}
  \includegraphics[scale=0.9]{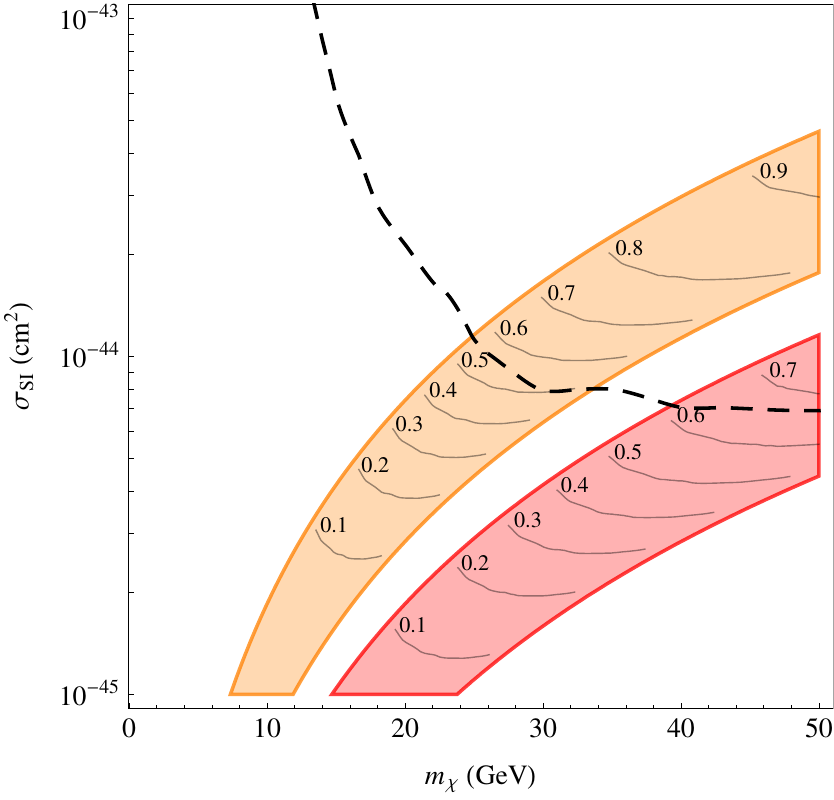}
\end{center}
\caption{Plot of $\sigma_{\rm SI}$ versus $m_\chi$ for complex scalar 
 dark matter.  The dashed black line depicts the constraint from XENON100. 
 The \{red, yellow\} bands indicate the range corresponding to $114~{\rm GeV} 
 < m_h < 145~{\rm GeV}$ for $n=\{1,2\}$, respectively, while the labeled 
 contours indicate the value of ${\rm BR}(h \rightarrow \chi\chi)$.}
\label{fig:xenon_scalar}
\end{figure}
\begin{figure}[t]
\begin{center}
  \includegraphics[scale=0.9]{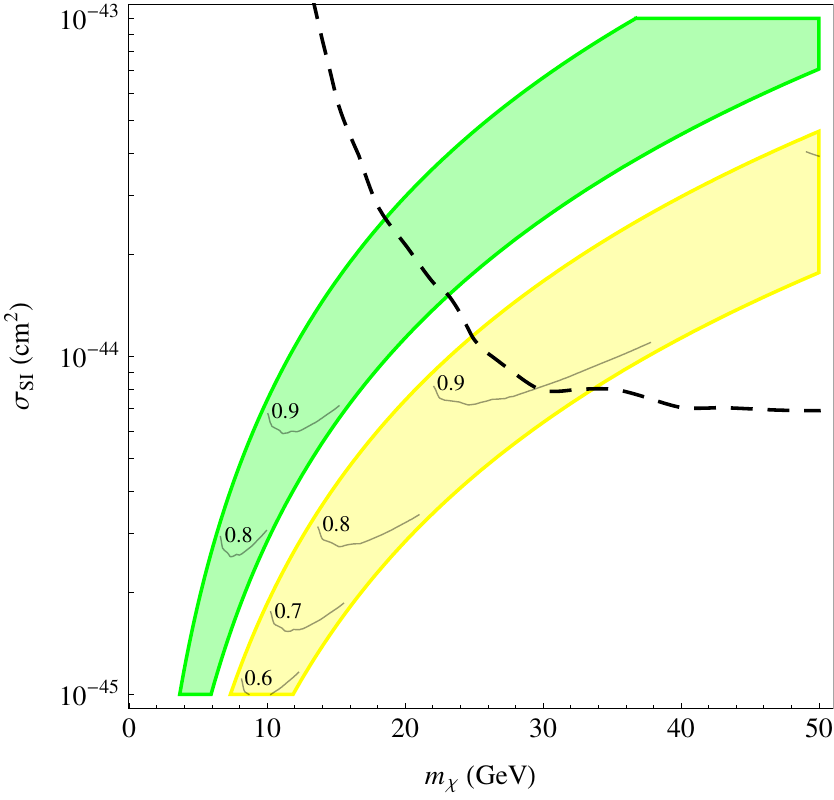}
\end{center}
\caption{Same as \Fig{fig:xenon_scalar} except for Dirac fermion dark 
 matter.  The \{yellow, green\} bands indicate the range corresponding 
 to $114~{\rm GeV} < m_h < 145~{\rm GeV}$ for $n=\{1,2\}$, respectively.}
\label{fig:xenon_fermion}
\end{figure}
The above result can now be compared with predictions for the invisible 
Higgs boson decay rate: $h \rightarrow \chi\chi$, where $\chi$ is neutral 
and stable component of the Higgs descendant.  In Figs.~\ref{fig:xenon_scalar} 
and \ref{fig:xenon_fermion}, we present contour plots of ${\rm BR}(h 
\rightarrow \chi\chi)$ in the $(m_\chi,\sigma_{\rm SI})$ planes for 
complex scalar and Dirac fermion dark matter, respectively.  Assuming 
that the dark matter-nucleon cross section saturates the lower bound 
in \Eq{eq:sigma}, each point in the $(m_\chi,\sigma_{\rm SI})$ plane 
corresponds to a fixed value of $m_h$ for a given $n$, and thus to a 
definite value of ${\rm BR}(h \rightarrow \chi\chi)$.  Using this, we 
have shown shaded bands corresponding to the Higgs boson mass range 
$114~{\rm GeV} < m_h < 145~{\rm GeV}$ for $n=1,2$ in each figure. 
Within these bands we have also drawn labeled contours, showing 
${\rm BR}(h \rightarrow \chi\chi)$.  Finally, the dashed black line 
in each figure indicates the most recent exclusion curve from the 
XENON100 dark matter detection experiment~\cite{Aprile:2011hi}. 
From \Fig{fig:xenon_scalar}, we see that if an invisible branching 
ratio of the Higgs boson greater than $\sim 70\%$ is measured, then 
complex scalar dark matter of this type will be excluded (up to 
astrophysical uncertainties associated with the dark matter detection 
constraint).  On the other hand, from \Fig{fig:xenon_fermion} it is 
clear that for Dirac fermion dark matter, a mostly invisibly decaying 
Higgs boson can easily be accommodated.

\section{Explicit Models}
\label{sec:models}

A number of prominent examples of Higgs descendants have been studied 
in the existing literature, albeit in varying contexts.  Perhaps the 
most obvious among these theories is singlet dark matter coupled via the 
so-called Higgs portal~\cite{Silveira:1985rk,Raidal:2011xk,McDonald:1993ex}:
\be
  {\cal L} = -\frac{\lambda}{4} |h|^2 |S|^2.
\ee
Because this model is rather phenomenological in nature, naturalness 
considerations are disregarded and a bare mass term for the gauge 
singlet, $m^2 |S|^2$, is simply not included.%
\footnote{Such a model can be made natural by the inclusion of 
 supersymmetry.}
The analyses of Secs.~\ref{sec:Higgs-prop} and \ref{sec:DM} apply 
more or less verbatim to this model, whose properties are correctly 
characterized by Figs.~\ref{fig:scalar} and \ref{fig:xenon_scalar}.

A less obvious example of a Higgs descendant is a fourth generation lepton, 
consisting of $L, E^c$ and $N^c$ with the interaction
\be
  {\cal L} = \lambda L h N^c + \kappa L h^\dagger E^c,
\label{eq:4gen}
\ee
where $L=(N,E)$ is a left-handed doublet.%
\footnote{The standard model gauge anomalies can be canceled by introducing, 
 e.g., a mirror lepton generation $(L^c,E,N)$ or a fourth generation 
 quark $(Q,U^c,D^c)$.}
The right-handed neutrino, $N^c$, is a gauge singlet, so in general it 
can have a Majorana mass term
\be
  {\cal L} = -\frac{m}{2} N^c N^c.
\label{eq:m}
\ee
However, as discussed earlier, the value of $m$ has no reason to be near 
the electroweak scale, so we expect it not to be very close to $v$.

If $m$ is sufficiently larger than $v$, then $N^c$ should be integrated 
out, yielding a higher dimension Majorana mass for $N$ given by $m_N 
= \lambda^2 v^2/m$.  Since $N$ acquires a mass solely from electroweak 
symmetry breaking, it is an example of a Higgs descendant, as expected from our general 
considerations.  Unfortunately, in the limit $m \gg v$ one expects 
$m_N < m_Z/2$, so this theory is disfavored from $Z$ pole measurements.

Alternatively, it may be that $m \ll v$, in which case $N^c$ remains 
in the spectrum and acquires a Dirac mass with $N$ such that $m_N 
= \lambda v$.  In this case $(N,N^c)$ is a Higgs descendant and 
Figs.~\ref{fig:fermion} and \ref{fig:xenon_fermion} pertain.  For $m_N 
> m_Z/2$ this theory trivially evades $Z$ pole constraints, but having 
$(N,N^c)$ dark matter may be difficult.  This is because \Fig{fig:xenon_fermion} 
implies $m_h \gtrsim 175~{\rm GeV}$, in which case there is no way 
to avoid the bound from standard model Higgs search~\cite{Higgs-LHC}, 
given that ${\rm BR}(h \rightarrow WW^*)$ cannot be suppressed; see 
\Fig{fig:fermion}.  The $(N,N^c)$ dark matter is possible if the 
XENON100 constraint is a factor of 2 weaker than that depicted in 
\Fig{fig:xenon_fermion}, due e.g.\ to astrophysical uncertainties. 
In this case, the direct detection constraint requires only $m_h 
\gtrsim 150~{\rm GeV}$, so that significant depletion of ${\rm BR}(h 
\rightarrow WW^*)$ is possible for $m_N < m_h/2$.

Let us now consider our final example.  Suppose there is a supersymmetric 
``hidden sector'' which has a $U(1)_X$ gauge field kinetically mixing 
with the hypercharge $U(1)_Y$ of the standard model~\cite{Holdom:1985ag}:
\be
  {\cal L} = \frac{\epsilon}{2} \int d^2 \theta\; {\cal W}_Y {\cal W}_X
  \supset \epsilon D_Y D_X.
\ee
The mixed $D$-term then produces a scalar potential of the form
\bea
  V &=& \frac{1}{2} \Bigl\{ g_X (|X|^2 - |X^c|^2) + \xi \Bigr\}^2,
\\
  \xi &=& \epsilon D_Y = -\frac{\epsilon g_Y}{2} |h|^2.
\eea
Here, $X$ and $X^c$ are hidden sector chiral superfields charged under 
$U(1)_X$.  For simplicity, we have taken the decoupling limit where 
$h$ is really the up-type Higgs boson at $\tan\beta = \infty$.  After 
electroweak symmetry breaking, $\xi$ acquires a VEV, triggering spontaneous 
symmetry breaking in the hidden sector proportional to the order parameter
\be
  \vev{X} = \sqrt{\frac{|\xi|}{g_X}},
\ee
where we have assumed $\epsilon > 0$.  Since $\vev{X}$ breaks the 
$U(1)_X$ gauge symmetry, the real and imaginary components of $X$ are 
eaten to become the longitudinal and radial modes of the massive vector 
supermultiplet, $V_X$.  The mass of this supermultiplet is given by 
$m_{V_X} = \sqrt{2 g_X |\xi|} \propto v$, so that $V_X$ is a Higgs 
descendant and our arguments apply to all its components.  (These 
components split in mass after supersymmetry breaking.)

\section{Discussion}
\label{sec:discuss}

In this paper we have analyzed a broad class of theories in which a new 
particle beyond the standard model, $\chi$, acquires its mass predominantly 
from the VEV of the Higgs field, $v$, rather than any intrinsic mass 
scale.  Such a particle, which we called a Higgs descendant, can arise 
naturally from any new sector whose intrinsic mass scales does not 
coincide with the electroweak scale.%
\footnote{The QCD scale is another scale in the standard model, which 
 is a priori independent of the electroweak symmetry breaking scale. 
 In principle, one may consider ``QCD descendant'' by coupling new 
 physics to a QCD condensate.}
Because the couplings of Higgs descendants are highly constrained, as 
given in \Eq{eq:hcoupling}, the physics is dictated essentially by the 
mass and spin of $\chi$.  As we have seen, both for light and heavy 
Higgs descendants there can be substantial modifications of the production 
and decay of the Higgs boson.  In the case where $\chi$ is also a dark 
matter particle, this class of theories predict a minimum spin-independent 
direct detection cross section that could be probed in experiments in 
the near future.

While our discussion has been limited to the case of a single Higgs 
doublet theory, our analysis can be straightforwardly extended to 
a two Higgs doublet model simply by expanding $m_\chi$ in a power 
series in $v_u$ and $v_d$.  However, this class of Higgs descendants 
becomes less predictive, due to a proliferation of free parameters 
coming from the multi-variate expansion of $m_\chi$ as well as Higgs 
boson mass and VEV mixing angles such as $\tan\alpha$ and $\tan\beta$. 
Our analysis, however, applies without modification in the decoupling 
regime.

Likewise, our discussion can also be extended to include supersymmetry. 
In many such theories, the interaction of \Eq{eq:hcoupling} is accompanied 
by a supersymmetric analog in which the Higgs boson is replaced by the 
Higgsino.  Hence, any given value of $m_\chi$ suggests a minimal coupling 
of the Higgsino to $\chi$ and its superpartner.  We leave an investigation 
of these possible Higgsino descendants for future work.

\begin{figure}[t]
\begin{center}
  \includegraphics[scale=0.9]{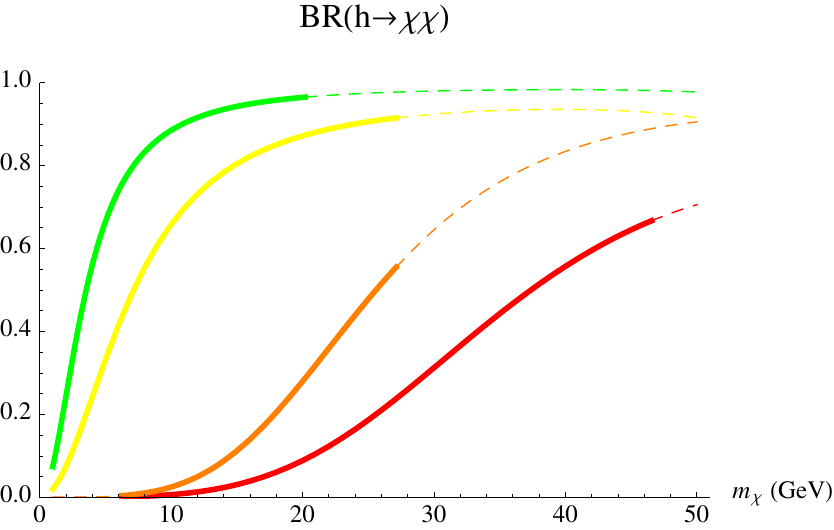}
\end{center}
\caption{Plot of ${\rm BR}(h \rightarrow \gamma\gamma)$ versus dark matter 
 mass $m_\chi$ for $m_h = 125$ GeV.  Here, \{red, orange\} corresponds to a complex scalar 
 with $n = \{1,2\}$ and \{yellow, green\} to a Dirac fermion with 
 $n = \{1,2\}$.  Dashed regions are excluded by XENON100.}
\label{fig:125}
\end{figure}
Lastly, let us briefly comment on the recently observed excesses seen 
at ATLAS and CMS in the di-gamma and $ZZ$ channels~\cite{Higgs-125}. 
Interpreted as a signal from the decay of the Higgs boson, this suggests 
a mass $m_h\simeq 125~{\rm GeV}$, although these experiments obviously 
cannot yet make definite statements about the detailed properties of such 
a Higgs particle.  As such, it is essential that further experimental 
analyses is applied to determine more precisely these properties.  Fixing 
$m_h = 125~{\rm GeV}$, any Higgs descendant theory has a parameter space 
which is even more restricted than that discussed in previous sections. 
In \Fig{fig:125} we have plotted the value of ${\rm BR}(h \rightarrow 
\chi\chi)$ for $n=1,2$ and for Dirac fermion and complex scalar $\chi$. 
The solid (dashed) portions of the curve correspond to regions in 
parameter space which are (dis)allowed by XENON100, if $\chi$ composes 
all of the dark matter.

\acknowledgments

We thank Tomer Volansky for collaboration at an early stage of this work 
and Joshua Ruderman for useful discussions.  This work was supported 
in part by the Director, Office of Science, Office of High Energy 
and Nuclear Physics, of the US Department of Energy under Contract 
DE-AC02-05CH11231, and in part by the National Science Foundation 
under grant PHY-0855653.

\end{document}